\documentclass[aip,amsmath,amssymb,reprint
 aip,
 amsmath,amssymb,
 reprint,%
]{revtex4-1}

\usepackage{graphicx}% Include figure files
\usepackage{dcolumn}% Align table columns on decimal point
\usepackage{bm}% bold math
\usepackage[utf8]{inputenc}
\usepackage[T1]{fontenc}
\usepackage{mathptmx}
\usepackage{etoolbox}
\usepackage{float}
\usepackage{siunitx}

\makeatletter
\def\@email#1#2{%
 \endgroup
 \patchcmd{\titleblock@produce}
  {\frontmatter@RRAPformat}
  {\frontmatter@RRAPformat{\produce@RRAP{*#1\href{mailto:#2}{#2}}}\frontmatter@RRAPformat}
  {}{}
}%
\makeatother
\begin{document}

\preprint{AIP/123-QED}

\title{In-vacuum surface flashover of SiN, AlN, and etched SiO$_2$ thin films at micrometre scales}
% Force line breaks with \\

% Force line breaks with \\
\author{Vijay~Kumar}
\affiliation{Sussex Centre for Quantum Technologies, University of Sussex, Brighton, BN1 9RH, U.K.}
\author{Martin~Siegele-Brown}
\affiliation{Sussex Centre for Quantum Technologies, University of Sussex, Brighton, BN1 9RH, U.K.}
\author{Matthew~Aylett}
\affiliation{Sussex Centre for Quantum Technologies, University of Sussex, Brighton, BN1 9RH, U.K.}
\author{Sebastian~Weidt}
\affiliation{Sussex Centre for Quantum Technologies, University of Sussex, Brighton, BN1 9RH, U.K.}
\affiliation{Universal Quantum Ltd, Brighton, BN1 6SB, U.K.}
\author{Winfried~Karl~Hensinger}
\affiliation{Sussex Centre for Quantum Technologies, University of Sussex, Brighton, BN1 9RH, U.K.}
\affiliation{Universal Quantum Ltd, Brighton, BN1 6SB, U.K.}
\email{w.k.hensinger@sussex.ac.uk}

\date{\today}% It is always \today, today,

\begin{abstract}
We investigate the surface flashover voltage threshold for SiO$_2$, SiN, and AlN thin films over micrometre scale lengths. Furthermore, we test the effects of different etching chemistries on SiO$_2$ layers. We find that there is little significant difference between untreated SiO$_2$ samples and those that have been etched with hydrogen fluoride or Transene AlPad Etch 639. SiN and AlN samples performed significantly better than all SiO$_2$ samples giving a 45\% increase in surface flashover voltage at a distance of \SI{5}{\micro\metre} with the difference increasing with electrode spacing. 
\end{abstract}

\maketitle

\begin{figure}[b!]
\centering

         \includegraphics[width=0.35\textwidth]{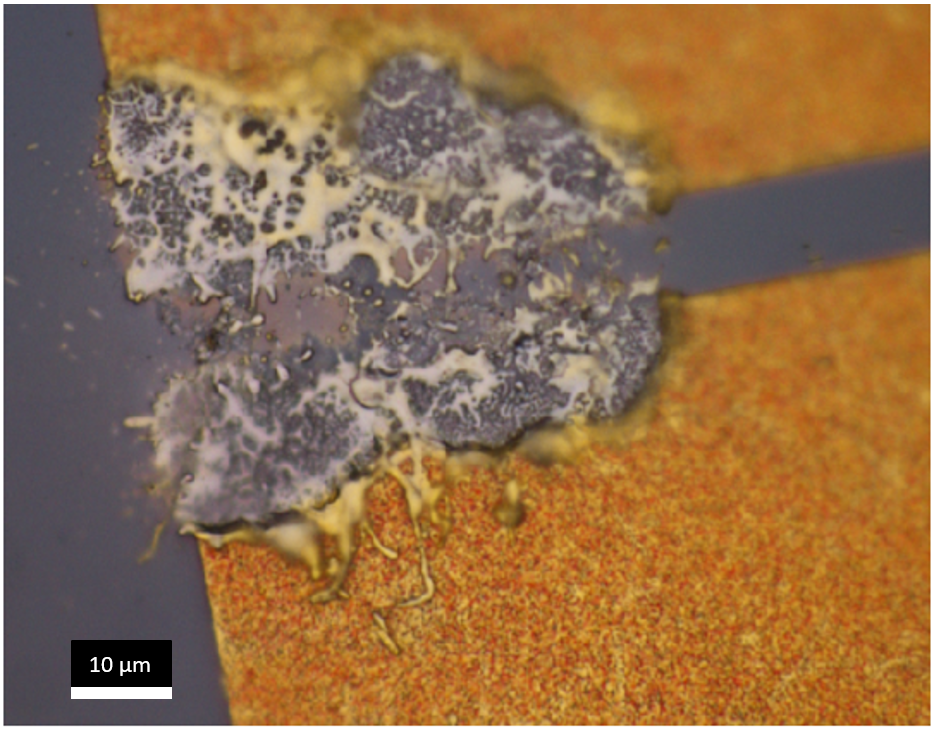}
         \caption{Sample damage after surface flashover. Damage is seen on the corners of the electrodes where the electric field is concentrated.
}
         \label{fig:electrode_damage}
\end{figure}

Understanding and being able to predict voltage discharge in a vacuum is crucial in many areas of technology. There are three general mechanisms in a vacuum that can cause an uncontrolled and unintended voltage discharge. These are: vacuum breakdown in which field emission of charge carriers causes a direct flashover to a nearby electrode\cite{vacuum}, bulk breakdown in which the electric field inside an insulator becomes strong enough to transition it into a conductor\cite{Bulk}, and finally surface flashover in which charge is carried over the surface of an insulator between two electrodes. Surface flashover seems to be a weak point in vacuum insulation with bulk and vacuum flashover voltage thresholds usually being much higher than that of the surface threshold. Vacuum isolation using dielectrics is used in many areas of electronics and technology\cite{circuitvacuum}, including pulsed power devices\cite{pulsed}, vacuum MEMS fabricated RF/microwave circuits\cite{rf}, and microfabricated ion and atom trapping microchips\cite{zak}. Moreover, in the fast growing field of ion trapping technology the application of high RF and DC voltages to MEMS fabricated chips is extremely common. The surface flashover threshold over micrometre scale lengths of dielectrics places an upper limit on the achievable trapping depth and secular frequency \cite{TRapD} for many microfabricated ion traps. Higher trap depths increase the ion lifetime \cite{hong2016guidelines, HightrapD} and higher radial secular frequencies reduce the Kerr coupling between the axial and radial modes during a qubit gate, which is a source of infidelity for quantum gates  \cite{nie2009theory}. In this case, surface flashover can cause damage as shown in Figure \ref{fig:electrode_damage} where surface flashover has occurred between two neighbouring electrodes which are separated by \SI{10}{\micro\metre} of SiO$_2$  rendering the device inoperable. As increasing the dielectric distance comes with challenges, finding material with a better surface flashover threshold can significantly improve microfabricated ion traps and other devices.
Furthermore, many of the early experiments that empirically influenced theoretical models of surface flashover investigated millimetre scale gaps, with little data on micrometre scale stacks that are present in ion trap devices. Therefore, it is important to understand the surface flashover limitations of materials that could be used in ion trapping devices and the effects of common etching processes over micrometre scales. 

While there is no consensus on the exact cause of surface flashover, it is agreed that an electron emission occurs either through field emission or thermal field emission. After this, the mechanism by which charge carriers travel along the surface of an insulator is debated with the most prevalent theory being that of the secondary electron emission avalanche (SEEA) \cite{SEEA}. In this mechanism, we assume a percentage of the emitted electrons will strike the insulator which has positive charges deposited on its surface producing additional electrons by secondary emission. Some of these secondary electrons will strike the insulator again creating a tertiary emission, and so forth. Additionally, the electron collision stimulates gas desorption from the insulator surface. As the desorbed gas reaches a certain pressure, high energy electrons in the electric field collide with gas molecules and cause collision ionisation\cite{plasma}. This creates a plasma extending across the anode-cathode gap further aiding flashover. Anderson et al. provide a detailed explanation of this theory along with experimental data \cite{anderson}. Pillai et al.\cite{eq} have found that the relationship between the DC voltage at which surface flashover occurs ($V_f$) and anode-cathode distance ($D$) is given as

\begin{equation}
    V_f=\sqrt{\frac{M_{cr}E_1v_0eD}{2\epsilon_0\gamma v_e \tan(\theta)}}.
\label{distance}
\end{equation}
$e$ and $\epsilon_0$ in equation \ref{distance} denote the charge of an electron and the permittivity of free space respectively. Other material dependent parameters include the amount of desorbed gas per unit area at the point of flashover $M_{cr}$, velocity of desorbed gas molecules $v_0$, and $\gamma$ the desorption probability of gas molecules. There is also dependence on the angle of impact $\theta$ of electrons colliding with the insulator, and the average electron drift velocity in the desorbed gas $v_e$ which is related to the electron impact energy $E_i$ by\cite{eq} $v_e = 5.94 \times 10^{5} \sqrt{E_i}$. In Pillai's work \cite{eq}, it is found that the surface charge density was correlated to the impact energy of electrons when striking the insulator surface, and indicates that the impact energy for sustained SEEA saturates near the lower critical energy $E_1$.

As there is little widely available data regarding the parameters $M_{cr}$, $E_1$, $v_0$, $\gamma$, and $v_e$ for different materials, we will group these together as a fitting parameter $\tau$, defining it as 
\begin{equation}
    \tau=\frac{M_{cr}E_1 v_0 }{\gamma v_e \tan(\theta)}.
\end{equation}
\noindent Previous literature has taken the same approach when studying the surface breakdown in different materials\cite{sterling}. 
From theory we expect the surface flashover threshold voltage and distance to flashover to scale as $V_f \propto \sqrt{D}$. However, the data for flashover over small distances on the micrometre scale is sparse with previous literature also indicating a dependence on the surface roughness and cleanliness of the insulator \cite{Rough}$^,$\cite{Grooves}$^,$\cite{rough2}. Sterling et al. \cite{sterling} report a 5\% difference between DC and RF surface flashover thresholds, a difference that was not statistically significant given the rest of the data. Consequently, we choose only to test DC flashover, as the experimental apparatus allows for more accurate DC measurements. 

\begin{figure}[b!]
\centering

         \includegraphics[width=0.3\textwidth]{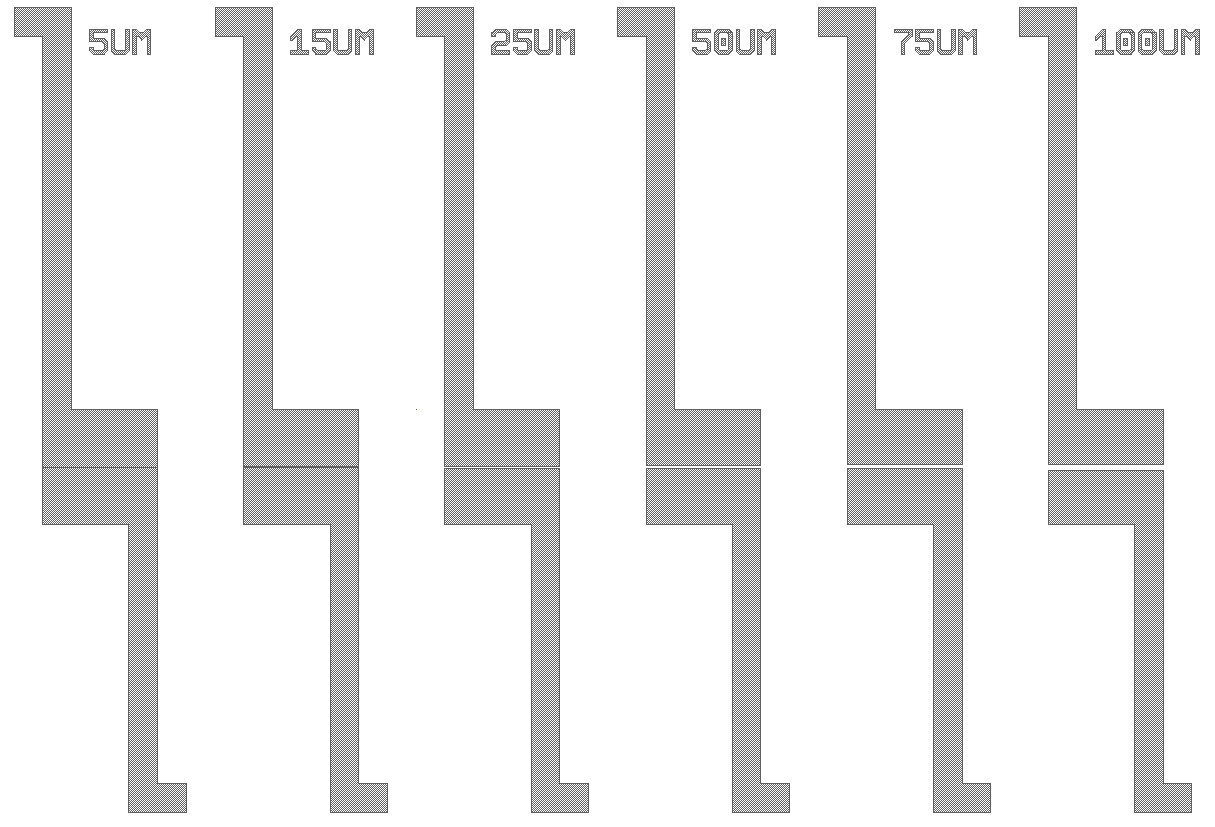}
         \caption{Electrode structure of surface flashover test chips. The distances to flashover are 5, 15, 25, 50, 75, and \SI{100}{\micro\metre}. 
}
         \label{fig:electrode_mask}
\end{figure}

To compare the surface flashover threshold for SiN, AlN, SiO$_2$, and the effects of etching chemistry, we fabricate a test chip on a fused silica wafer. SiN, AlN, and SiO$_2$ are chosen for testing as they are commonly used materials in MEMS fabrication and can be easily integrated into ion trap designs. Fused silica is used instead of a semiconductive wafer to avoid bulk breakdown down to the substrate. A \SI{0.5}{\micro\metre} thick test film is then deposited. SiO$_2$ and SiN samples were deposited via PECVD and AlN samples were deposited via sputtering. Next a Au electrode is deposited and patterned through lift off lithography. Lift off lithography was chosen so as not to affect the surface chemistry or topology of the test film with an etching process. To test the effect of commonly used etching chemistry on the flashover voltage threshold, some SiO$_2$ samples were exposed for \SI{5}{s} in a buffered hydrogen fluoride solution, or Transene AlPad Etch 639 (a common etchant used to etch SiO$_2$ which does not affect Al pads). \SI{5}{s} is chosen to provide a minimal etch of the material to maintain the flashover distance while changing the surface chemistry and topography using the etching solution. The anode-cathode electrode distance $D$ varies from 5 to \SI{100}{\micro\metre} on a chip.

\begin{figure}[t!]
\centering

         \includegraphics[width=0.45\textwidth]{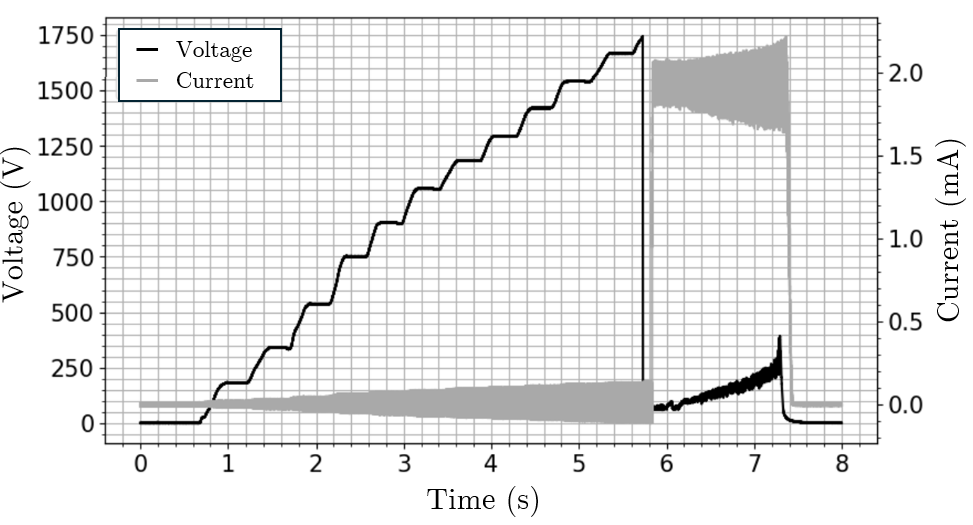}
         \caption{Raw IV data collected from a sample under test. The voltage is increased until surface flashover is detected by a spike in measured current. Surface flashover occurs above approximately \SI{1.5}{kV}. 
}
         \label{fig:one_run}
\end{figure}

The test chips are mounted on a PCB in a vacuum chamber at a pressure of $\approx 1 \times 10^{-6}$mbar. A variable DC power supply was used to apply a voltage to the electrodes with the current and voltage being monitored with two digital multimeters that are triggered together and are capable of 1M samples per second recording frequency (Keithley DAQ6510). The voltage applied to the chip was then increased from 0 until surface flashover occurred. With this current and voltage monitoring we are able to spot instantaneous flashovers through a spike in current and match the timestamp to the voltage that was applied at the time of flashover. A sample of data is shown in figure \ref{fig:one_run} where surface flashover occurs at above \SI{1.5}{kV}. This method was validated by filming the test and spotting the visible flash and roughly matching it to the time recorded by the multimeter.

\begin{figure}[t!]
\centering

         \includegraphics[width=0.42\textwidth]{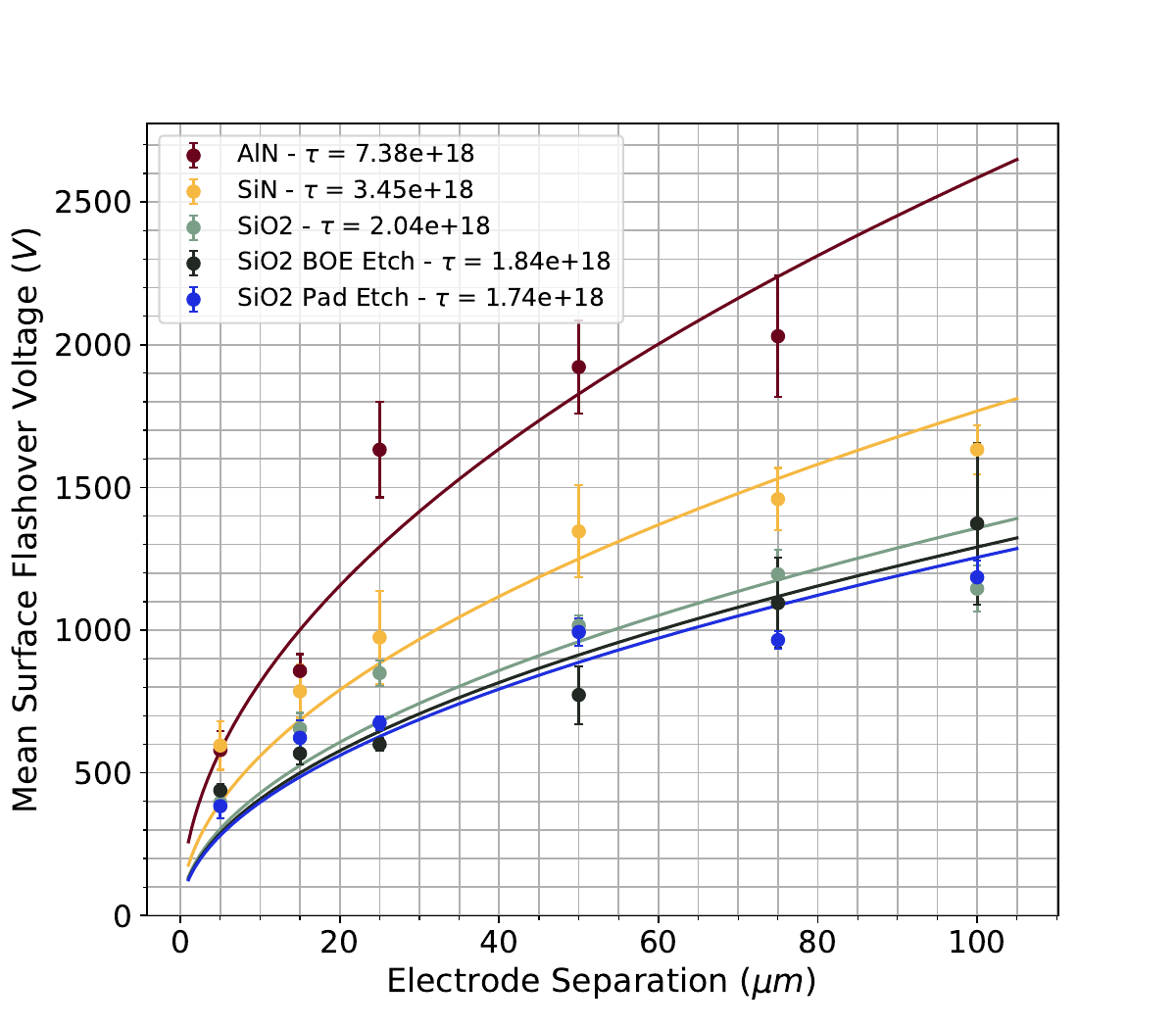}
         \caption{Surface flashover as function of electrode separation for different dielectrics (SiO$_2$, SiN, AlN) response from samples. The measurement data is fitted to equation \ref{distance} using a non-linear least squares regression method. The units of $\tau$ are in V/m$^2$. Data for AlN above 100$\mu$m was not recorded as the applied voltage caused breakdown of the mounting PCB before breakdown of the device under test.}  
         \label{fig:grpah}
\end{figure}

Damage caused by surface flashover on the samples is shown in figure \ref{fig:electrode_damage}. Flashes usually occur at the edges of the electrodes which is where we expect the electric field to be concentrated. Results of the experiment are shown in figure \ref{fig:grpah}. The AlN and SiN samples show significantly higher surface breakdown thresholds compared to SiO$_2$ samples. Secondly, etching chemistry seemed to have a very limited effect on surface flashover. From the data we can see over a small gap of \SI{5}{\micro\metre}, on the length scales found in ion trap microchips, there is a significant increase of around \SI{200}{V} in surface flashover threshold when comparing SiN and AlN samples to SiO$_2$. Furthermore, AlN is shown to be the most resilient to surface flashover with the surface flashover threshold exceeding \SI{2}{kV} at \SI{75}{\micro\metre}.

In conclusion we investigate the surface flashover voltage threshold of SiN, AlN, SiO$_2$ along with looking into the effect of different etching chemistries. We focus on micrometre scale lengths and materials used in MEMS devices that could be used in vacuum. It was found that exposing SiO$_2$ samples to buffered hydrogen fluoride and Transene AlPad Etch 639 had little significant effect on surface flashover voltage threshold. SiN and AlN samples performed much better than SiO$_2$ samples with the difference in performance increasing with anode-cathode distance. AlN samples performed best with the highest breakdown voltage at over 2kV, however, at smaller anode-cathode gaps the difference between SiN and AlN is negligible. In the context of MEMS fabrication of ion traps where gap sizes between electrodes and ground can vary between 3 to \SI{20}{\micro\metre}, voltages over \SI{150}{V} are regularly applied. The surface flashover threshold can be increased by hundreds of voltages by switching from commonly used SiO$_2$ to AlN or SiN for dielectric insulation. As the achieved trap depth is proportional to the peak RF voltage squared \cite{hong2016guidelines}, an increase of 45\% at a distance of \SI{5}{\micro\metre} translates to a $\approx$ 110\% increase in trap depth, potentially leading to longer ion lifetimes and increased robustness of microfabricated ion trap devices.

\section*{Data availability statement}
%The data that support the findings of this study are openly available at the following URL/DOI: xxx.
The data that support the findings of this study are openly available in the supplementary information. 

\section*{Acknowledgments}

Work was carried out at a number of facilities including the Center of MicroNanoTechnology (CMi) at École Polytechnique Fédérale de Lausanne (EPFL). This work was supported by the U.K. Engineering and Physical Sciences Research Council via the EPSRC Hub in Quantum Computing and Simulation (EP/T001062/1), the U.K. Quantum Technology hub for Networked Quantum Information Technologies (No. EP/M013243/1), the European Commission's Horizon-2020 Flagship on Quantum Technologies Project No. 820314 (MicroQC), the U.S. Army Research Office under Contract No. W911NF-14-2-0106 and Contract No. W911NF-21-1-0240, the Office of Naval Research under Agreement No. N62909-19-1-2116, and the University of Sussex.

\section*{References}
\bibliography{Ref}% Produces the bibliography via BibTeX.

\begin{thebibliography}{18}%
\makeatletter
\providecommand \@ifxundefined [1]{%
 \@ifx{#1\undefined}
}%
\providecommand \@ifnum [1]{%
 \ifnum #1\expandafter \@firstoftwo
 \else \expandafter \@secondoftwo
 \fi
}%
\providecommand \@ifx [1]{%
 \ifx #1\expandafter \@firstoftwo
 \else \expandafter \@secondoftwo
 \fi
}%
\providecommand \natexlab [1]{#1}%
\providecommand \enquote  [1]{``#1''}%
\providecommand \bibnamefont  [1]{#1}%
\providecommand \bibfnamefont [1]{#1}%
\providecommand \citenamefont [1]{#1}%
\providecommand \href@noop [0]{\@secondoftwo}%
\providecommand \href [0]{\begingroup \@sanitize@url \@href}%
\providecommand \@href[1]{\@@startlink{#1}\@@href}%
\providecommand \@@href[1]{\endgroup#1\@@endlink}%
\providecommand \@sanitize@url [0]{\catcode `\\12\catcode `\$12\catcode `\&12\catcode `\#12\catcode `\^12\catcode `\_12\catcode `\%12\relax}%
\providecommand \@@startlink[1]{}%
\providecommand \@@endlink[0]{}%
\providecommand \url  [0]{\begingroup\@sanitize@url \@url }%
\providecommand \@url [1]{\endgroup\@href {#1}{\urlprefix }}%
\providecommand \urlprefix  [0]{URL }%
\providecommand \Eprint [0]{\href }%
\providecommand \doibase [0]{http://dx.doi.org/}%
\providecommand \selectlanguage [0]{\@gobble}%
\providecommand \bibinfo  [0]{\@secondoftwo}%
\providecommand \bibfield  [0]{\@secondoftwo}%
\providecommand \translation [1]{[#1]}%
\providecommand \BibitemOpen [0]{}%
\providecommand \bibitemStop [0]{}%
\providecommand \bibitemNoStop [0]{.\EOS\space}%
\providecommand \EOS [0]{\spacefactor3000\relax}%
\providecommand \BibitemShut  [1]{\csname bibitem#1\endcsname}%
\let\auto@bib@innerbib\@empty
%</preamble>
\bibitem [{\citenamefont {Li}\ \emph {et~al.}(2017)\citenamefont {Li}, \citenamefont {Geng}, \citenamefont {Liu},\ and\ \citenamefont {Wang}}]{vacuum}%
  \BibitemOpen
  \bibfield  {author} {\bibinfo {author} {\bibfnamefont {S.}~\bibnamefont {Li}}, \bibinfo {author} {\bibfnamefont {Y.}~\bibnamefont {Geng}}, \bibinfo {author} {\bibfnamefont {Z.}~\bibnamefont {Liu}}, \ and\ \bibinfo {author} {\bibfnamefont {J.}~\bibnamefont {Wang}},\ }\bibfield  {title} {\enquote {\bibinfo {title} {A breakdown mechanism transition with increasing vacuum gaps},}\ }\href {\doibase 10.1109/TDEI.2017.006482} {\bibfield  {journal} {\bibinfo  {journal} {IEEE Transactions on Dielectrics and Electrical Insulation}\ }\textbf {\bibinfo {volume} {24}},\ \bibinfo {pages} {3340--3346} (\bibinfo {year} {2017})}\BibitemShut {NoStop}%
\bibitem [{\citenamefont {Li}, \citenamefont {Tung},\ and\ \citenamefont {Pey}(2008)}]{Bulk}%
  \BibitemOpen
  \bibfield  {author} {\bibinfo {author} {\bibfnamefont {X.}~\bibnamefont {Li}}, \bibinfo {author} {\bibfnamefont {C.}~\bibnamefont {Tung}}, \ and\ \bibinfo {author} {\bibfnamefont {K.}~\bibnamefont {Pey}},\ }\bibfield  {title} {\enquote {\bibinfo {title} {The nature of dielectric breakdown},}\ }\href@noop {} {\bibfield  {journal} {\bibinfo  {journal} {Applied Physics Letters}\ }\textbf {\bibinfo {volume} {93}} (\bibinfo {year} {2008})}\BibitemShut {NoStop}%
\bibitem [{\citenamefont {König}(2012)}]{circuitvacuum}%
  \BibitemOpen
  \bibfield  {author} {\bibinfo {author} {\bibfnamefont {D.}~\bibnamefont {König}},\ }\bibfield  {title} {\enquote {\bibinfo {title} {The role of vacuum in circuit breaker technology},}\ }in\ \href {\doibase 10.1109/DEIV.2012.6412366} {\emph {\bibinfo {booktitle} {2012 25th International Symposium on Discharges and Electrical Insulation in Vacuum (ISDEIV)}}}\ (\bibinfo {year} {2012})\ pp.\ \bibinfo {pages} {A1--A14}\BibitemShut {NoStop}%
\bibitem [{\citenamefont {Qi}\ \emph {et~al.}(2017)\citenamefont {Qi}, \citenamefont {Gao}, \citenamefont {Sun},\ and\ \citenamefont {Li}}]{pulsed}%
  \BibitemOpen
  \bibfield  {author} {\bibinfo {author} {\bibfnamefont {B.}~\bibnamefont {Qi}}, \bibinfo {author} {\bibfnamefont {C.}~\bibnamefont {Gao}}, \bibinfo {author} {\bibfnamefont {Z.}~\bibnamefont {Sun}}, \ and\ \bibinfo {author} {\bibfnamefont {C.}~\bibnamefont {Li}},\ }\bibfield  {title} {\enquote {\bibinfo {title} {Surface charge accumulation of solid insulator under nanosecond pulse in vacuum: 3d distribution features and mechanism},}\ }\href {\doibase 10.1088/1361-6463/aa8ce5} {\bibfield  {journal} {\bibinfo  {journal} {Journal of Physics D: Applied Physics}\ }\textbf {\bibinfo {volume} {50}},\ \bibinfo {pages} {465603} (\bibinfo {year} {2017})}\BibitemShut {NoStop}%
\bibitem [{\citenamefont {Patton}\ and\ \citenamefont {Zabinski}(2005)}]{rf}%
  \BibitemOpen
  \bibfield  {author} {\bibinfo {author} {\bibfnamefont {S.}~\bibnamefont {Patton}}\ and\ \bibinfo {author} {\bibfnamefont {J.}~\bibnamefont {Zabinski}},\ }\bibfield  {title} {\enquote {\bibinfo {title} {Fundamental studies of au contacts in mems rf switches},}\ }\href@noop {} {\bibfield  {journal} {\bibinfo  {journal} {Tribology Letters}\ }\textbf {\bibinfo {volume} {18}},\ \bibinfo {pages} {215--230} (\bibinfo {year} {2005})}\BibitemShut {NoStop}%
\bibitem [{\citenamefont {Romaszko}\ \emph {et~al.}(2020)\citenamefont {Romaszko}, \citenamefont {Hong}, \citenamefont {Siegele}, \citenamefont {Puddy}, \citenamefont {Lebrun-Gallagher}, \citenamefont {Weidt},\ and\ \citenamefont {Hensinger}}]{zak}%
  \BibitemOpen
  \bibfield  {author} {\bibinfo {author} {\bibfnamefont {Z.~D.}\ \bibnamefont {Romaszko}}, \bibinfo {author} {\bibfnamefont {S.}~\bibnamefont {Hong}}, \bibinfo {author} {\bibfnamefont {M.}~\bibnamefont {Siegele}}, \bibinfo {author} {\bibfnamefont {R.~K.}\ \bibnamefont {Puddy}}, \bibinfo {author} {\bibfnamefont {F.~R.}\ \bibnamefont {Lebrun-Gallagher}}, \bibinfo {author} {\bibfnamefont {S.}~\bibnamefont {Weidt}}, \ and\ \bibinfo {author} {\bibfnamefont {W.~K.}\ \bibnamefont {Hensinger}},\ }\bibfield  {title} {\enquote {\bibinfo {title} {Engineering of microfabricated ion traps and integration of advanced on-chip features},}\ }\href@noop {} {\bibfield  {journal} {\bibinfo  {journal} {Nature Reviews Physics}\ }\textbf {\bibinfo {volume} {2}},\ \bibinfo {pages} {285--299} (\bibinfo {year} {2020})}\BibitemShut {NoStop}%
\bibitem [{\citenamefont {Auchter}\ \emph {et~al.}(2022)\citenamefont {Auchter}, \citenamefont {Axline}, \citenamefont {Decaroli}, \citenamefont {Valentini}, \citenamefont {Purwin}, \citenamefont {Oswald}, \citenamefont {Matt}, \citenamefont {Aschauer}, \citenamefont {Colombe}, \citenamefont {Holz}, \citenamefont {Monz}, \citenamefont {Blatt}, \citenamefont {Schindler}, \citenamefont {Rössler},\ and\ \citenamefont {Home}}]{TRapD}%
  \BibitemOpen
  \bibfield  {author} {\bibinfo {author} {\bibfnamefont {S.}~\bibnamefont {Auchter}}, \bibinfo {author} {\bibfnamefont {C.}~\bibnamefont {Axline}}, \bibinfo {author} {\bibfnamefont {C.}~\bibnamefont {Decaroli}}, \bibinfo {author} {\bibfnamefont {M.}~\bibnamefont {Valentini}}, \bibinfo {author} {\bibfnamefont {L.}~\bibnamefont {Purwin}}, \bibinfo {author} {\bibfnamefont {R.}~\bibnamefont {Oswald}}, \bibinfo {author} {\bibfnamefont {R.}~\bibnamefont {Matt}}, \bibinfo {author} {\bibfnamefont {E.}~\bibnamefont {Aschauer}}, \bibinfo {author} {\bibfnamefont {Y.}~\bibnamefont {Colombe}}, \bibinfo {author} {\bibfnamefont {P.}~\bibnamefont {Holz}}, \bibinfo {author} {\bibfnamefont {T.}~\bibnamefont {Monz}}, \bibinfo {author} {\bibfnamefont {R.}~\bibnamefont {Blatt}}, \bibinfo {author} {\bibfnamefont {P.}~\bibnamefont {Schindler}}, \bibinfo {author} {\bibfnamefont {C.}~\bibnamefont {Rössler}}, \ and\ \bibinfo {author} {\bibfnamefont {J.}~\bibnamefont {Home}},\ }\bibfield  {title} {\enquote {\bibinfo {title}
  {Industrially microfabricated ion trap with 1 ev trap depth},}\ }\href {\doibase 10.1088/2058-9565/ac7072} {\bibfield  {journal} {\bibinfo  {journal} {Quantum Science and Technology}\ }\textbf {\bibinfo {volume} {7}},\ \bibinfo {pages} {035015} (\bibinfo {year} {2022})}\BibitemShut {NoStop}%
\bibitem [{\citenamefont {Hong}\ \emph {et~al.}(2016)\citenamefont {Hong}, \citenamefont {Lee}, \citenamefont {Cheon}, \citenamefont {Kim},\ and\ \citenamefont {Cho}}]{hong2016guidelines}%
  \BibitemOpen
  \bibfield  {author} {\bibinfo {author} {\bibfnamefont {S.}~\bibnamefont {Hong}}, \bibinfo {author} {\bibfnamefont {M.}~\bibnamefont {Lee}}, \bibinfo {author} {\bibfnamefont {H.}~\bibnamefont {Cheon}}, \bibinfo {author} {\bibfnamefont {T.}~\bibnamefont {Kim}}, \ and\ \bibinfo {author} {\bibfnamefont {D.-i.~D.}\ \bibnamefont {Cho}},\ }\bibfield  {title} {\enquote {\bibinfo {title} {Guidelines for designing surface ion traps using the boundary element method},}\ }\href@noop {} {\bibfield  {journal} {\bibinfo  {journal} {Sensors}\ }\textbf {\bibinfo {volume} {16}},\ \bibinfo {pages} {616} (\bibinfo {year} {2016})}\BibitemShut {NoStop}%
\bibitem [{\citenamefont {Suleimen}\ \emph {et~al.}(2024)\citenamefont {Suleimen}, \citenamefont {Podlesnyy}, \citenamefont {Akopyan}, \citenamefont {Sterligov}, \citenamefont {Lakhmanskaya}, \citenamefont {Anikin}, \citenamefont {Matveev},\ and\ \citenamefont {Lakhmanskiy}}]{HightrapD}%
  \BibitemOpen
  \bibfield  {author} {\bibinfo {author} {\bibfnamefont {Y.}~\bibnamefont {Suleimen}}, \bibinfo {author} {\bibfnamefont {A.}~\bibnamefont {Podlesnyy}}, \bibinfo {author} {\bibfnamefont {L.~A.}\ \bibnamefont {Akopyan}}, \bibinfo {author} {\bibfnamefont {N.}~\bibnamefont {Sterligov}}, \bibinfo {author} {\bibfnamefont {O.}~\bibnamefont {Lakhmanskaya}}, \bibinfo {author} {\bibfnamefont {E.}~\bibnamefont {Anikin}}, \bibinfo {author} {\bibfnamefont {A.}~\bibnamefont {Matveev}}, \ and\ \bibinfo {author} {\bibfnamefont {K.}~\bibnamefont {Lakhmanskiy}},\ }\bibfield  {title} {\enquote {\bibinfo {title} {Surface trap with adjustable ion couplings for scalable and parallel gates},}\ }\href {\doibase 10.1103/PhysRevA.109.022605} {\bibfield  {journal} {\bibinfo  {journal} {Phys. Rev. A}\ }\textbf {\bibinfo {volume} {109}},\ \bibinfo {pages} {022605} (\bibinfo {year} {2024})}\BibitemShut {NoStop}%
\bibitem [{\citenamefont {Nie}, \citenamefont {Roos},\ and\ \citenamefont {James}(2009)}]{nie2009theory}%
  \BibitemOpen
  \bibfield  {author} {\bibinfo {author} {\bibfnamefont {X.~R.}\ \bibnamefont {Nie}}, \bibinfo {author} {\bibfnamefont {C.~F.}\ \bibnamefont {Roos}}, \ and\ \bibinfo {author} {\bibfnamefont {D.~F.}\ \bibnamefont {James}},\ }\bibfield  {title} {\enquote {\bibinfo {title} {Theory of cross phase modulation for the vibrational modes of trapped ions},}\ }\href@noop {} {\bibfield  {journal} {\bibinfo  {journal} {Physics Letters A}\ }\textbf {\bibinfo {volume} {373}},\ \bibinfo {pages} {422--425} (\bibinfo {year} {2009})}\BibitemShut {NoStop}%
\bibitem [{\citenamefont {Miller}(1989)}]{SEEA}%
  \BibitemOpen
  \bibfield  {author} {\bibinfo {author} {\bibfnamefont {H.}~\bibnamefont {Miller}},\ }\bibfield  {title} {\enquote {\bibinfo {title} {Surface flashover of insulators},}\ }\href {\doibase 10.1109/14.42158} {\bibfield  {journal} {\bibinfo  {journal} {IEEE Transactions on Electrical Insulation}\ }\textbf {\bibinfo {volume} {24}},\ \bibinfo {pages} {765--786} (\bibinfo {year} {1989})}\BibitemShut {NoStop}%
\bibitem [{\citenamefont {Li}\ \emph {et~al.}(2023)\citenamefont {Li}, \citenamefont {Liu}, \citenamefont {Ohki}, \citenamefont {Chen}, \citenamefont {Gao},\ and\ \citenamefont {Li}}]{plasma}%
  \BibitemOpen
  \bibfield  {author} {\bibinfo {author} {\bibfnamefont {Z.}~\bibnamefont {Li}}, \bibinfo {author} {\bibfnamefont {J.}~\bibnamefont {Liu}}, \bibinfo {author} {\bibfnamefont {Y.}~\bibnamefont {Ohki}}, \bibinfo {author} {\bibfnamefont {G.}~\bibnamefont {Chen}}, \bibinfo {author} {\bibfnamefont {H.}~\bibnamefont {Gao}}, \ and\ \bibinfo {author} {\bibfnamefont {S.}~\bibnamefont {Li}},\ }\bibfield  {title} {\enquote {\bibinfo {title} {Surface flashover in 50 years: Theoretical models and competing mechanisms},}\ }\href {\doibase https://doi.org/10.1049/hve2.12340} {\bibfield  {journal} {\bibinfo  {journal} {High Voltage}\ }\textbf {\bibinfo {volume} {8}},\ \bibinfo {pages} {853--877} (\bibinfo {year} {2023})}\BibitemShut {NoStop}%
\bibitem [{\citenamefont {Anderson}\ and\ \citenamefont {Brainard}(1980)}]{anderson}%
  \BibitemOpen
  \bibfield  {author} {\bibinfo {author} {\bibfnamefont {R.~A.}\ \bibnamefont {Anderson}}\ and\ \bibinfo {author} {\bibfnamefont {J.~P.}\ \bibnamefont {Brainard}},\ }\bibfield  {title} {\enquote {\bibinfo {title} {Mechanism of pulsed surface flashover involving electron‐stimulated desorption},}\ }\href {\doibase 10.1063/1.327839} {\bibfield  {journal} {\bibinfo  {journal} {Journal of Applied Physics}\ }\textbf {\bibinfo {volume} {51}},\ \bibinfo {pages} {1414--1421} (\bibinfo {year} {1980})}\BibitemShut {NoStop}%
\bibitem [{\citenamefont {Pillai}\ and\ \citenamefont {Hackam}(1982)}]{eq}%
  \BibitemOpen
  \bibfield  {author} {\bibinfo {author} {\bibfnamefont {A.~S.}\ \bibnamefont {Pillai}}\ and\ \bibinfo {author} {\bibfnamefont {R.}~\bibnamefont {Hackam}},\ }\bibfield  {title} {\enquote {\bibinfo {title} {Surface flashover of solid dielectric in vacuum},}\ }\href {\doibase 10.1063/1.331037} {\bibfield  {journal} {\bibinfo  {journal} {Journal of Applied Physics}\ }\textbf {\bibinfo {volume} {53}},\ \bibinfo {pages} {2983--2987} (\bibinfo {year} {1982})}\BibitemShut {NoStop}%
\bibitem [{\citenamefont {Sterling}\ \emph {et~al.}(2013)\citenamefont {Sterling}, \citenamefont {Hughes}, \citenamefont {Mellor},\ and\ \citenamefont {Hensinger}}]{sterling}%
  \BibitemOpen
  \bibfield  {author} {\bibinfo {author} {\bibfnamefont {R.~C.}\ \bibnamefont {Sterling}}, \bibinfo {author} {\bibfnamefont {M.~D.}\ \bibnamefont {Hughes}}, \bibinfo {author} {\bibfnamefont {C.~J.}\ \bibnamefont {Mellor}}, \ and\ \bibinfo {author} {\bibfnamefont {W.~K.}\ \bibnamefont {Hensinger}},\ }\bibfield  {title} {\enquote {\bibinfo {title} {Increased surface flashover voltage in microfabricated devices},}\ }\href {\doibase 10.1063/1.4824012} {\bibfield  {journal} {\bibinfo  {journal} {Applied Physics Letters}\ }\textbf {\bibinfo {volume} {103}},\ \bibinfo {pages} {143504} (\bibinfo {year} {2013})}\BibitemShut {NoStop}%
\bibitem [{\citenamefont {Naruse}\ \emph {et~al.}(2015)\citenamefont {Naruse}, \citenamefont {Saito}, \citenamefont {Sakaki},\ and\ \citenamefont {Yamamoto}}]{Rough}%
  \BibitemOpen
  \bibfield  {author} {\bibinfo {author} {\bibfnamefont {H.}~\bibnamefont {Naruse}}, \bibinfo {author} {\bibfnamefont {H.}~\bibnamefont {Saito}}, \bibinfo {author} {\bibfnamefont {M.}~\bibnamefont {Sakaki}}, \ and\ \bibinfo {author} {\bibfnamefont {O.}~\bibnamefont {Yamamoto}},\ }\bibfield  {title} {\enquote {\bibinfo {title} {Flashover mechanisms of bridged vacuum gaps based on cathode electric field measurement},}\ }\href {\doibase 10.1109/TDEI.2014.004566} {\bibfield  {journal} {\bibinfo  {journal} {IEEE Transactions on Dielectrics and Electrical Insulation}\ }\textbf {\bibinfo {volume} {22}},\ \bibinfo {pages} {597--603} (\bibinfo {year} {2015})}\BibitemShut {NoStop}%
\bibitem [{\citenamefont {Zhang}\ \emph {et~al.}(2020)\citenamefont {Zhang}, \citenamefont {Sun}, \citenamefont {Mu}, \citenamefont {Song}, \citenamefont {Xue},\ and\ \citenamefont {Zhang}}]{Grooves}%
  \BibitemOpen
  \bibfield  {author} {\bibinfo {author} {\bibfnamefont {S.}~\bibnamefont {Zhang}}, \bibinfo {author} {\bibfnamefont {G.-Y.}\ \bibnamefont {Sun}}, \bibinfo {author} {\bibfnamefont {H.-B.}\ \bibnamefont {Mu}}, \bibinfo {author} {\bibfnamefont {B.-P.}\ \bibnamefont {Song}}, \bibinfo {author} {\bibfnamefont {J.}~\bibnamefont {Xue}}, \ and\ \bibinfo {author} {\bibfnamefont {G.-J.}\ \bibnamefont {Zhang}},\ }\bibfield  {title} {\enquote {\bibinfo {title} {Modelling vacuum flashover mitigation with complex surface microstructure: mechanism and application},}\ }\href {\doibase https://doi.org/10.1049/hve.2019.0363} {\bibfield  {journal} {\bibinfo  {journal} {High Voltage}\ }\textbf {\bibinfo {volume} {5}},\ \bibinfo {pages} {110--121} (\bibinfo {year} {2020})}\BibitemShut {NoStop}%
\bibitem [{\citenamefont {Guo}\ \emph {et~al.}(2019)\citenamefont {Guo}, \citenamefont {Sun}, \citenamefont {Zhang}, \citenamefont {Xue}, \citenamefont {Zhou}, \citenamefont {Song}, \citenamefont {Mu},\ and\ \citenamefont {Zhang}}]{rough2}%
  \BibitemOpen
  \bibfield  {author} {\bibinfo {author} {\bibfnamefont {B.-H.}\ \bibnamefont {Guo}}, \bibinfo {author} {\bibfnamefont {G.-Y.}\ \bibnamefont {Sun}}, \bibinfo {author} {\bibfnamefont {S.}~\bibnamefont {Zhang}}, \bibinfo {author} {\bibfnamefont {J.-Y.}\ \bibnamefont {Xue}}, \bibinfo {author} {\bibfnamefont {R.-D.}\ \bibnamefont {Zhou}}, \bibinfo {author} {\bibfnamefont {B.-P.}\ \bibnamefont {Song}}, \bibinfo {author} {\bibfnamefont {H.-B.}\ \bibnamefont {Mu}}, \ and\ \bibinfo {author} {\bibfnamefont {G.-J.}\ \bibnamefont {Zhang}},\ }\bibfield  {title} {\enquote {\bibinfo {title} {Mechanism of vacuum flashover on surface roughness},}\ }\href {\doibase 10.1088/1361-6463/ab05a0} {\bibfield  {journal} {\bibinfo  {journal} {Journal of Physics D: Applied Physics}\ }\textbf {\bibinfo {volume} {52}},\ \bibinfo {pages} {215301} (\bibinfo {year} {2019})}\BibitemShut {NoStop}%
\end{thebibliography}%

\onecolumngrid
\section*{Supplementary information}

The five tables below list the breakdown data for all samples measured.

% Requires: \usepackage{array}
\begin{table}[H]
    \centering
    \caption{SiO$_2$ breakdown data}
    \begin{tabular}{|c|c|c|c|c|c|}
        \hline
        Gap [\SI{}{\micro\metre}] & Sample 1 [V]  & Sample 2 [V]  & Sample 3 [V]  & Sample 4 [V]  & Sample 5 [V]  \\ \hline
        5	&	417	&	339	&		&	431	&	394	 \\ \hline
        15	&	480	&	662	&	701	&	777	&		 \\ \hline
        25	&		&	824	&	718	&	910	&	947	 \\ \hline
        50	&	960	&	987	&	1100	&		&		 \\ \hline
        75	&		&	1365	&	1032	&	1013	&	1373	 \\ \hline
        100	&		&		&	1260	&		&	1030	 \\ \hline
    \end{tabular}
    
\caption{SiO$_2$ BOE Etch breakdown data}
    \begin{tabular}{|c|c|c|c|c|c|}
        \hline
        Gap [\SI{}{\micro\metre}] & Sample 1 [V]  & Sample 2 [V]  & Sample 3 [V]  & Sample 4 [V]  \\ \hline
        5	&	419	&	466	&	495	&	374			 \\ \hline
        15	&	508	&	546	&	698	&	520			 \\ \hline
        25	&		&	632	&	568	&				 \\ \hline
        50	&	662	&	1124	&	671	&	636			 \\ \hline
        75	&	1448	&	1055	&	782	&				 \\ \hline
        100	&	1774	&		&	973	&				 \\ \hline

    \end{tabular}
    
\caption{SiO$_2$ Pad Etch breakdown data}
    \begin{tabular}{|c|c|c|c|c|c|}
        \hline
        Gap [\SI{}{\micro\metre}] & Sample 1 [V]  & Sample 2 [V]  & Sample 3 [V]  \\ \hline
5	&	323	&	338	&	490					 \\ \hline
15	&	491	&	620	&	757					 \\ \hline
25	&	642	&	709	&						 \\ \hline
50	&	880	&	1021	&	1080					 \\ \hline
75	&		&	921	&	1009					 \\ \hline
100	&	1047	&	1232	&	1276					 \\ \hline

    \end{tabular}
    
\caption{AlN breakdown data}
    \begin{tabular}{|c|c|c|c|c|c|}
        \hline
        Gap [\SI{}{\micro\metre}] & Sample 1 [V]  & Sample 2 [V]  & Sample 3 [V]  & Sample 4 [V]  \\ \hline
5	&	420	&	794	&	553	&	553			 \\ \hline
15	&	940	&	774	&		&				 \\ \hline
25	&		&	1396	&	1868	&				 \\ \hline
50	&	1532	&		&	2202	&	2032			 \\ \hline
75	&	1515	&		&	2215	&	2359			 \\ \hline

    \end{tabular}
    
\caption{SiN breakdown data}
    \begin{tabular}{|c|c|c|c|c|c|c|}
        \hline
        Gap [\SI{}{\micro\metre}] & Sample 1 [V]  & Sample 2 [V]  & Sample 3 [V]  & Sample 4 [V]  & Sample 5 [V]  & Sample 6 [V]  \\ \hline
5	&	731	&	810	&		&	303	&	448	&	689	 \\ \hline
15	&	925	&		&		&	560	&		&	872	 \\ \hline
25	&		&		&	1283	&	601	&		&	1040	 \\ \hline
50	&	1319	&		&	1154	&	992	&	2034	&	1231	 \\ \hline
75	&		&	1139	&	1455	&		&	1489	&		 \\ \hline
100	&		&		&		&		&	1512	&	1754	 \\ \hline

    \end{tabular}
    
    \label{tab:placeholder_label}
\end{table}

\end{document}